# The Cydoc smart patient intake form accelerates medical note writing

Angela Hemesath MSE[1,2], Kenyon Wright[1,2] BS, Matthew Michael Draelos, MD, PhD[3], Rachel Lea Draelos, MD, PhD[1]

[1]Cydoc Corporation, Durham, NC 27705
[2]Duke University School of Medicine, Durham, NC 27710
[3]University of North Carolina, Chapel Hill, NC 27514

Corresponding Author: Rachel Lea Draelos, MD, PhD. rachel [dot] draelos [at] cydoc.ai

**Abstract**

Purpose: This study evaluates the effect of Cydoc software tools on medical note time-to-completion and quality.

Methods: Medical students were recruited by email to participate in a video encounter with a standardized patient for three scenarios: writing a note from scratch (control), writing a note with the Cydoc educational tool, and writing a note with the Cydoc intake form. Notes were subsequently anonymized and rated by a resident physician across four quality measures. Note time-to-completion was analyzed using a one-way ANOVA with post-hoc Bonferroni correction, while note quality scores were compared using a Wilcoxon paired signed rank test.

Results: Eighteen medical students participated in the study. The average note time-to-completion, which included the patient interview and note writing, was $17 \pm 7.0$ minutes from scratch, $18 \pm 8.0$ minutes with the educational tool, and $5.7 \pm 3.0$ minutes with the intake form. Using the Cydoc intake form was significantly faster than writing from scratch ($p = 0.0001$) or using the educational tool ($p = 8 \times 10^{-5}$). Notes written with Cydoc tools had higher note comprehensiveness (3.24 > 3.06), pertinent positives (3.47 > 2.94), and pertinent negatives (3.47 > 2.67), although this trend did not reach statistical significance.

Conclusions: Using the Cydoc smart patient intake form accelerated note writing by 2.98× while maintaining note quality. The Cydoc smart patient intake form has the potential to streamline clinical documentation and save clinicians' time. Future work is needed to evaluate Cydoc tools in an in-person outpatient setting with practicing clinician users.

## 1. INTRODUCTION

Medical note writing often consumes multiple hours of a clinician's workday, with physicians spending about one-third of their time on patient documentation.[1] Many physicians also spend time working on notes from home in the evenings. Unfortunately, increased time spent after hours on electronic medical records is associated with burnout and lower work-life satisfaction. [2,3]

There is thus significant interest in developing software to accelerate note writing. Previous studies have shown that health information technology can improve medical note-writing tasks by up to 12-fold [4-6] with smart tools in surgery increasing note content accuracy [7] and smart tools in chronic disease addressing more deficiencies in care.[8]

Cydoc is a health tech startup that aims to alleviate documentation burden through smart tools targeting outpatient primary care clinics and emergency departments. Cydoc has developed a proprietary artificial intelligence system that interviews patients using clinical reasoning, then transforms semi-structured patient data into the text of a medical note. This AI system has been integrated into two smart tools: the Cydoc educational tool and the Cydoc smart patient intake form. Both tools are cloud-based and launched through a web browser for use on any desktop or mobile device. The smart patient intake form is available at www.cydoc.ai.

The Cydoc educational tool provides medical students with a streamlined interface for each section of a note. The interface helps the medical student remember what questions to ask a patient during an interview, including through detailed chief-complaint-specific questionnaires in the history of present illness (HPI) section. The goal of this tool is to help medical students learn to take a comprehensive yet focused history.

The Cydoc smart patient intake form is intended to replace existing paper and digital patient intake forms. A patient fills out the Cydoc intake form before their visit, for example at home or in a waiting room. Then, Cydoc's AI engine transforms the intake form data into the text of a medical note, which is available to the clinician at the start of an encounter. The clinician can then refine the existing Cydoc note instead of writing a new note from scratch. The goal of this tool is to save time through two mechanisms: first, by streamlining the patient encounter itself through automated collection of key information beforehand, and second, by reducing the amount of time required to write high-quality notes.

The primary objective of this study was to determine whether the Cydoc educational tool and Cydoc intake form affect note time-to-completion. The secondary objective was to determine whether Cydoc smart tools affect note quality, as rated by a resident physician.

## 2. METHODS

This study was performed in accordance with the Health Insurance Portability and Accountability Act (HIPAA) and adhered to the principles of the Declaration of Helsinki. This study was deemed exempt by the Pearl Institutional Review Board (IRB).

.

## 2.1 Subjects

Eighteen medical students from year one through year four were recruited via email between December 2022 and January 2023. Medical students were all enrollees in good standing at Duke University (17 students) and the University of Virginia (1 student).

Participants engaged in a video call with a standardized patient (A.H., K.W.), in which each participant interviewed the standardized patient once for each of three separate clinical scenarios: back pain, abdominal pain, and diabetes. Details of the clinical scenarios were defined and provided to the standardized patient beforehand. A summary of each clinical scenario is available in the Supplementary Material.

Medical students used a different note writing method for each scenario:
- Writing a note from scratch, meaning no note writing software was used beyond standard word processing software;
- "Educational tool," meaning the Cydoc educational tool was used interactively by the participant during the encounter; and
- "Intake form," meaning the note generated by the Cydoc smart patient intake form was provided to the participant at the start of the encounter, simulating a situation where the standardized patient filled out the smart form in the waiting room. The student could thus initialize their note with the note generated by the smart patient intake form.

The order of clinical scenarios and order of note writing methods were randomly assigned and unique for each participant. The set permutations used for this random assignment process are shown in Supplementary Table 1.

Each note was required to include the following sections: history of present illness (HPI), patient history (past medical history, surgical history, medications, allergies, social history, family history), and review of systems (ROS). Physical exam was excluded due to the virtual nature of the encounter. Assessment/plan was also excluded, as this study focused on the history-taking aspects of medical note writing.

## 2.2 Outcome Measure: Note Time-to-Completion

The main goal of this study was to determine the effect of Cydoc smart tools on note time-to-completion. Each participant recorded the start time and stop time of each scenario. When writing a note from scratch or with Cydoc's educational tool, the start time was defined as the start of the verbal interviewing of the standardized patient. When the educational tool was used, this start time also coincided with the start of interactions with the educational tool. When writing a note with the Cydoc smart patient intake form, the start time was defined as the time that the participant began reading the generated note before interviewing the patient. The stop time for all note writing methods was defined as the time the participant completed their note. Total time was calculated as start time subtracted from stop time. Thus, for all scenarios, total time encompassed the entire patient interview as well as the entire note-writing process.

## 2.3 Outcome Measure: Note Quality

The secondary goal of this study was to determine whether Cydoc smart tools affected note quality, as an ideal smart tool would accelerate note writing without compromising note quality.

Each medical note was stripped of any information related to the participant or note writing method used. Notes were then processed into a consistent format of Times New Roman, size 12 font, with no bold, italics, or underline formatting. After de-identification, the notes were presented to a resident physician (M.M.D.) for rating. The resident physician had no involvement in the recruitment of participants or standardized patient encounters. Each note was rated on four characteristics across a 1 (poor) to 5 (excellent) scale:

- Clarity: the note was clear and easy to read.
- Comprehensiveness: the note included all medical information needed.
- Pertinent positives: all pertinent positives were included.
- Pertinent negatives: all pertinent negatives were included.

## 2.4 Data Collection

Google Docs was used to collect medical notes and survey data from participants. Each participant was provided with their own Google Doc, containing instructions for each stage of the process according to the unique note writing method and clinical scenario ordering assigned to that participant. Participants wrote their notes directly into the Google Doc and were permitted to use any built-in word processing features available. The Cydoc intake form scenario included the generated note text directly within the Google Doc. An example Google Doc is shown in the Supplementary Material. No personally identifiable information about participants was recorded within the Google Doc.

## 2.5 Statistical Analysis

Note time-to-completion was analyzed with a one-way ANOVA, with $p = 0.05$. A Bonferroni correction was applied to ANOVA results to correct for multiple testing. Note quality scores were analyzed with a Wilcoxon paired signed rank test. All statistical analysis was done using Microsoft Excel (version 16.70).

## 3. RESULTS

### 3.1 Note Time-to-Completion: the Cydoc smart patient intake form triples note-writing speed

The average time to write a note was $17 \pm 7.0$ minutes (mean ± standard deviation) from scratch and $18 \pm 8.0$ minutes for the Cydoc educational tool. Seventeen participants took between 1 and 13 minutes (average of $5.7 \pm 3$ minutes) to complete the intake form, while one participant required 29 minutes. This was 8 standard deviations above the mean. In the following analysis, this outlier has been excluded. Additionally, two participants were unable to generate a note from the Cydoc educational tool due to an unforeseen issue with a user interface element which was corrected after study completion. As a result, the time and note from these two encounters were excluded from statistical analysis. To view the raw timing data as well as the results of statistical analysis with the outlier included, see the Supplementary Material; all conclusions remain the same.

A one-way, paired ANOVA revealed a significant difference in note time-to-completion between groups (p-value = $2.04 \times 10^{-5}$). Post-hoc Bonferroni confirmed a significant difference in note time-to-completion between writing from scratch and the Cydoc intake form (p-value = $1.20 \times 10^{-4}$), as well as between the Cydoc educational tool and Cydoc intake form (p-value = $8.14 \times 10^{-5}$). No significant difference was found in note time-to-completion between writing from scratch and Cydoc's educational tool. Timing results are illustrated in Figure 1. A summary of note time-to-completion is illustrated in the Supplementary Material.

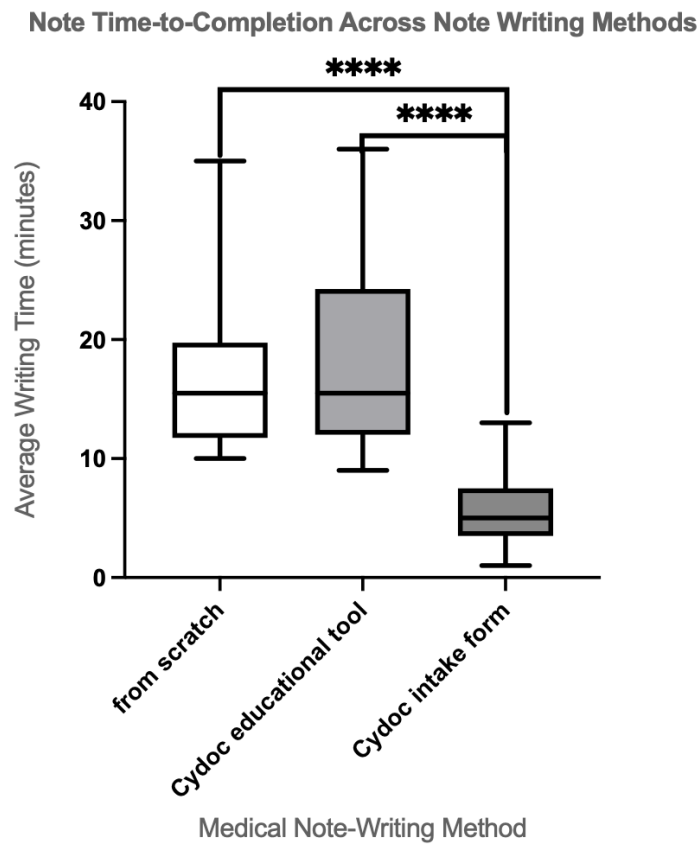

*Figure 1.* Note time-to-completion for three different note writing methods: from scratch (white box), Cydoc educational tool (light grey box), and Cydoc intake form (dark grey box). Significant differences are indicated with asterisks (asterisks between from scratch and Cydoc intake form and double asterisks between Cydoc educational tool and Cydoc intake form). * = p ≤ 0.05, ** = p ≤ 0.01, *** = p ≤ 0.001, **** = p ≤ 0.0001.

## 3.2 Note Quality: The Cydoc smart tools maintain note quality

Notes written with Cydoc tools had higher note comprehensiveness (3.24 > 3.06), pertinent positives (3.47 > 2.94), and pertinent negatives (3.47 > 2.67), although this trend did not reach statistical significance (Figure 2). Statistical analysis of note quality scores is available in the Supplementary Materials.

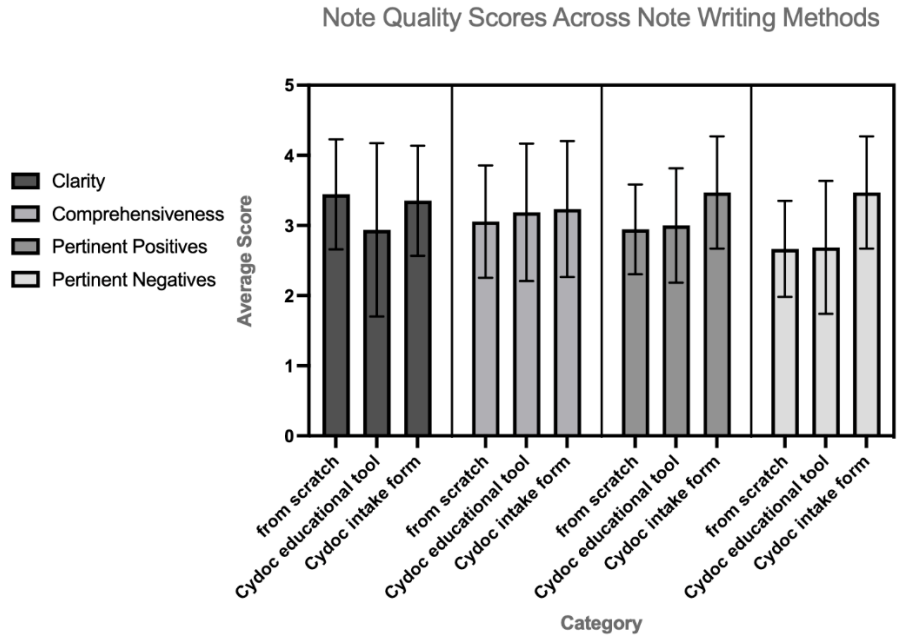

*Figure 2.* Note quality scores assigned to medical notes by an independent physician for notes written from scratch, with the Cydoc educational tool, and with the Cydoc intake form. Positive and negative standard deviation are shown with vertical lines.

## 4. DISCUSSION

The Cydoc intake form resulted in statistically significant, 3x faster note completion than writing a note from scratch, while the Cydoc educational tool did not affect note writing time. Both the Cydoc intake form and the Cydoc educational tool had higher average comprehensibility, pertinent positives, and pertinent negatives than writing a note from scratch, although these trends did not reach statistical significance.

### 4.1 Limitations and Future Work

The participants in this study were medical students, who likely take longer to write notes than experienced clinicians. Medical students were chosen as the study population to enable simultaneously studying the Cydoc educational tool, which targets students, alongside the Cydoc intake form, which is intended to benefit any clinician working in an outpatient setting. It is likely that the observed 3x time savings would differ if this study were repeated with experienced physicians as participants. However, even a smaller time savings of 1.5 – 2x has the potential to save multiple hours per week.

In the intake form scenario, the observed time savings could arise from a faster patient encounter, due to needing to ask fewer questions of the patient, and/or from a reduced need for typing due to the availability of an automatically generated draft note. We empirically observed that both factors contributed to the time saving effect, but we did not measure these two phases separately, due to the complexity that would entail for participants who type while talking to the patient. In future work it would be interesting to leverage keyboard tracking technology to better understand the factors that contribute to time saving aspects of the intake form.

This study relied upon an interaction with a standardized patient in a video conference setting. Standardized patient interactions can be more coherent and organized than some real-world patient interactions, and telehealth may be more conducive to new technology adoption.

We included three different note-writing scenarios in this study, but did not include the following two scenarios: one in which partial information on a patient was provided beforehand, to simulate browsing through a patient's electronic medical record before a visit, and another in which the user could leverage macros or dot phrases during note writing. These scenarios were not included due to the expense of creating a realistic EMR simulator.

In future work, we aim to repeat this study in an in-person clinical setting, with experienced clinicians as the users and real patients participating in the visits.

## 5. CONCLUSION

The Cydoc smart patient intake form accelerates note writing time and may be a useful tool to streamline the clinical documentation process.

**CONFLICT OF INTEREST**

R.L.D. is the CEO & Founder of Cydoc, the company that funded this research project. A.H. and K.W. are paid interns at Cydoc. M.M.D. has a financial interest in Cydoc.

**References**

1. Joukes E, Abu-Hanna A, Cornet R, de Keizer NF. Time Spent on Dedicated Patient Care and Documentation Tasks Before and After the Introduction of a Structured and Standardized Electronic Health Record. *Appl Clin Inform*. 2018;9(1):46-53. doi:10.1055/s-0037-1615747
2. Robertson SL, Robinson MD, Reid A. Electronic Health Record Effects on Work-Life Balance and Burnout Within the I3 Population Collaborative. *J Grad Med Educ*. 2017;9(4):479-484. doi:10.4300/JGME-D-16-00123.1
3. Bates DW, Gawande AA. Improving safety with information technology. *The New England Journal of Medicine*. https://www.nejm.org/doi/full/10.1056/NEJMsa020847. Published June 19, 2003. Accessed January 11, 2023.
4. Hunt DL, Haynes BR, Hanna SE. Effects of computer-based clinical decision support systems on physician performance and patient outcomes. *JAMA*. https://jamanetwork.com/journals/jama/article-abstract/188081?casa_token=Wr67eECBqUcAAAAA%3Ab4Gr2pm0ENeFEyMz7jPY5WPEr3xagWSFVqNqmcZGP6a7pmOVeVuyRqZYQFSTcBKZdMusID_HhQiIpQ. Published October 21, 1998. Accessed January 11, 2023.
5. Chertow GM, Lee J, Kuperman GJ. Guided medication dosing for inpatients with renal insufficiency. *JAMA*. https://jamanetwork.com/journals/jama/article-abstract/194455. Published December 12, 2001. Accessed January 11, 2023.
6. Chaudhry B, Wang J, Wu S, Maglione M, Mojica W, Roth E, et al. Systematic review: impact of health information technology on quality, efficiency, and costs of medical care. *Ann Intern Med* 2006 May 16;144(10):742–52.
7. Ghania Y, Thakrarb R, Kosugea D, Bate P. 'smart' electronic operation notes in surgery: An innovative way to improve patient care. *International Journal of Surgery*. https://www.sciencedirect.com/science/article/pii/S1743919113010935. Published November 14, 2013. Accessed January 11, 2023.
8. Schnipper JL, Linder JA, Palchuk MB, et al. Effects of documentation-based decision support on chronic disease management. *Am J Manag Care*. 2010;16(12 Suppl HIT):SP72-81.

# SUPPLEMENTARY MATERIAL

*Supplementary Table 1*. Randomization of note writing method and clinical scenario for the 18 participants in this study. CS = clinical scenario. CS1 = abdominal pain, CS2 = diabetes, CS3 = back pain. Note writing methods are "from scratch" (only word processing software used), "ed tool" (Cydoc educational tool used interactively), and "intake form" (Cydoc smart intake form generated note provided at the start of the encounter).

| Participants 1 – 6 | Participants 7 – 12 | Participants 13 – 18 |
|---|---|---|
| CS1 – from scratch<br>CS2 – ed tool<br>CS3 – intake form | CS2 – ed tool<br>CS3 – intake form<br>CS1 – from scratch | CS3 – intake form<br>CS1 – from scratch<br>CS2 – ed tool |
| CS1 – from scratch<br>CS2 – intake form<br>CS3 – ed tool | CS2 – intake form<br>CS3 – ed tool<br>CS1 – from scratch | CS3 – ed tool<br>CS1 – from scratch<br>CS2 – intake form |
| CS1 – intake form<br>CS2 – from scratch<br>CS3 – ed tool | CS2 – from scratch<br>CS3 – ed tool<br>CS1 – intake form | CS3 – ed tool<br>CS1 – intake form<br>CS2 – from scratch |
| CS1 – intake form<br>CS2 – ed tool<br>CS3 – from scratch | CS2 – ed tool<br>CS3 – from scratch<br>CS1 – intake form | CS3 – from scratch<br>CS1 – intake form<br>CS2 – ed tool |
| CS1 – ed tool<br>CS2 – intake form<br>CS3 – from scratch | CS2 – intake form<br>CS3 – from scratch<br>CS1 – ed tool | CS3 – from scratch<br>CS1 – ed tool<br>CS2 – intake form |
| CS1 – ed tool<br>CS2 – from scratch<br>CS3 – intake form | CS2 – from scratch<br>CS3 – intake form<br>CS1 – ed tool | CS3 – intake form<br>CS1 – ed tool<br>CS2 – from scratch |

*Supplementary Figure 1*. Standardized patient(SP) scenarios. Each participant interviewed the SP on all three scenarios. SPs followed the outline of information for each clinical scenario.

**Information for Standardized Patient with Chief Complaint "Back Pain"**

Cause (**do not share with student**): Muscle strain

Introductory Sentence: Ms. Smith is a 32-year-old woman presenting to a family physician's office with "back pain." She is a new patient in this health system.

| | |
|---|---|
| Chief Complaint | Back Pain |
| HPI | Back pain started 1 week ago after moving here from Virginia and carrying heavy boxes/furniture. <br> Back pain is on the right side in the middle. <br> Pain is constant, 5/10, dull, aching. <br> Better with lying flat in bed and hot packs. Worse with moving around and twisting. <br> Had pain like this two years ago after trying rock climbing. |
| Review of Systems | +back pain, cough <br> Negative for everything else |
| Medical History | Iron deficiency anemia since 2014 <br> Asthma since 2003 |
| Surgical History | Appendectomy in 2016 <br> Wisdom tooth removal in 2008 |
| Medications | Daily vitamin <br> Oral contraceptives |
| Allergies | Cats: coughing/sneezing |
| Social History | Current smoker, smoked 1 pack/day for 10 years <br> 2 glasses of wine per week <br> Lives alone <br> Works as real estate agent |
| Family History | Type II diabetes in mother <br> Heart failure in grandfather |

**Information for the Standardized Patient with Chief Complaint "Abdominal Pain"**

Cause (**do not share with student**): Viral gastroenteritis

Introductory Sentence: Ms. Jones is a 56-year-old woman presenting to the emergency department with "abdominal pain." She is a new patient in this health system.

| | |
|---|---|
| Chief Complaint | Abdominal Pain |
| HPI | Abdominal pain started 2 days ago along with other symptoms listed in ROS. Never had pain like this before. Crampy, sharp pain. 7/10 right now, 9/10 at its worst. The pain is diffuse throughout the abdomen. The pain is better with rest. Worse with eating. Patient is scared of needing surgery again. |
| Review of Systems | +abdominal pain, diarrhea (nonbloody), fever to 101 with thermometer, chills, nausea, vomiting (nonbloody) Negative for everything else. |
| Medical History | Hypertension Hyperlipidemia Myocardial infarction in 2017, stent placed Osteoarthritis |
| Surgical History | Cholecystectomy for cholelithiasis in 2004 |
| Medications | Amlodipine for hypertension Atorvastatin for hyperlipidemia Aspirin for osteoarthritis |
| Allergies | No allergies |
| Social History | Never smoker Drinks 5 beers a week Sexually active with husband Works as an elementary school teacher |
| Family History | Ulcerative colitis in sister Colon cancer in father, uncle, and grandfather |

**Information for the Standardized Patient with "Diabetes Outpatient Visit"**

Cause (**do not share with student**): Uncontrolled type 2 diabetes with hyperglycemia symptoms

| Chief Complaint | Diabetes Outpatient Visit |
|---|---|
| HPI | Diagnosed with Type II diabetes 5 years ago. Symptoms listed in ROS. Was last seen for her diabetes 2 years ago by her PCP (who has since retired). Once a day in the morning she measures blood sugar at home with a finger stick. Usually in the 200s or 300s. Doesn't write it down. Does not take insulin. Takes metformin and sometimes misses doses. For breakfast has cereal. For lunch and dinner, has meat and vegetables from her garden. For snacks has crackers and cheese. Exercises once per week by walking around the block. |
| Review of Systems | + polydipsia, polyuria, blurry vision, numbness in feet, tingling in feet. Negative for everything else. |
| Medical History | Type II diabetes<br>Obesity<br>Hypertension<br>Anxiety<br>Chronic kidney disease |
| Surgical History | C-sections in 2001 and 2006 |
| Medications | Metformin for diabetes<br>Alprazolam for anxiety |
| Allergies | Pollen: itchy eyes |
| Social History | Smoked 1 pack per day for 20 years<br>No alcohol use, no recreational drugs<br>Lives with her husband and their 16-year-old son<br>Works as an accountant |
| Family History | Type 2 diabetes in mother, father, and sister<br>Breast cancer in aunt |

*Supplementary Figure 2.* Example of the blank document given to participants to fill out as the study progressed. Participants were instructed to fill out the document themselves by recording their own start and stop time where specified. Survey questions after each section were included in the document.

**Overall Instructions**

- Fill out all sections highlighted in yellow.
- In this study you will be asked to write medical notes. Your goal should be to write notes that you would be willing to submit to an attending physician in a clinical setting.
- In your notes, ONLY include the following sections: History of Present Illness, Past Medical History, Surgical History, Medications, Allergies, Social History, Family History, and Review of Systems.
- Do NOT include a Physical Exam or an Assessment/Plan.

**Chief Complaint #1: Diabetes Outpatient Visit**

**Writing a Note with the Patient Intake Form**

This task is to receive a Patient Intake Form and use it as the starting point for your note.

The standardized patient will share a link to a separate Google Doc. This separate Google Doc contains the Patient Intake Form corresponding to a new chief complaint.

Record the time that you open the Google Doc and begin reading the Patient Intake Form:

| START TIME |
|---|
|  |

In a clinical setting, would you make edits to the Patient Intake Form before sending it to your attending as your note? Type your answer "YES" or "NO" into the box below:

| ARE EDITS NEEDED? |
|---|
|  |

If you answered "YES" to the previous question, paste the Patient Intake Form again in this box, and make the desired edits until you determine the note is complete.

Patient Intake Form, WITH EDITS:

|  |
|---|
|  |

Record the time that you complete your note:

| STOP TIME |
|---|
|  |

**Chief Complaint #2: Back Pain**

**Writing a Note with the Cydoc Note Writer**

This task is to use the Cydoc note writer while you interview the standardized patient.

Navigate to www.cydoc.ai in a web browser. The standardized patient will provide you with login credentials and an MFA code.

Record the time that you begin interviewing the standardized patient while using the Cydoc note writer:

| START TIME |
|---|
|  |

When you have finished interviewing the standardized patient, copy and paste the Cydoc Generated Note here:

PASTE CYDOC GENERATED NOTE HERE, WITHOUT ANY CHANGES:

|  |
|---|

In a clinical setting, would you make edits to the Cydoc Generated Note before sending it to your attending as your note? Type your answer "YES" or "NO" into the box below:

| ARE EDITS NEEDED? |
|---|
|  |

If you answered "YES" to the previous question, paste the Cydoc Generated Note again in this box, and make the desired edits until you determine the note is complete. Do NOT include the physical exam or assessment/plan sections.

CYDOC GENERATED NOTE, WITH EDITS:

|  |
|---|

Record the time that you complete your note:

| STOP TIME |
|---|
|  |

**Chief Complaint #3: Abdominal Pain**

**Writing a Note From Scratch**

This task is to interview the standardized patient and write a note starting from the blank template below.

Record the time that you begin interviewing the standardized patient:

| START TIME |
|---|
|  |

WRITE YOUR NOTE IN THIS BOX:

|  |
|---|

Record the time that you complete your note:

| STOP TIME |
|---|
|  |

**Survey Questions**

*[After completing the study, participants were then asked 6 survey questions for internal product development use.]*

END OF DOCUMENT

*Supplementary Table 2.* Average and standard deviation of timing data when writing from scratch, using the note-writing tool, and using the patient intake form. The table includes average timing data with all results, and timing data after outliers were removed.

| | Average note-writing time | | | |
|---|---|---|---|---|
| | From Scratch | Cydoc Note-Writer | Patient Intake Form | |
| average time | 17.0 | 18.4 | 7.0 | All subjects included |
| standard deviation | 7.0 | 8.0 | 6.2 | |
| average time | 17.0 | 18.4 | 5.7 | After outlier removal |
| standard deviation | 7.0 | 8.0 | 6.2 | |

*Supplementary Table 3*. One-way, paired ANOVA comparing note time-to-completion for different note writing methods.

**ANOVA: Single Factor**

SUMMARY

| Groups | Count | Sum | Average | Variance |
|---|---|---|---|---|
| from scratch | 18 | 306 | 17 | 48.8 |
| Cydoc educational tool | 16 | 294 | 18.375 | 63.3 |
| Cydoc intake form | 18 | 126 | 7 | 38.7 |

ANOVA

| Source of Variation | SS | df | MS | F | P-value | F crit |
|---|---|---|---|---|---|---|
| Between Groups | 1350.2 | 2.0 | 675.1 | 13.6 | 2.04E-05 | 3.2 |
| Within Groups | 2437.8 | 49.0 | 49.8 | | | |
| Total | 3787.9 | 51.0 | | | | |

Post-Hoc Bonferroni

| Groups | previous significant p-value | post-hoc Bonferroni p-value | between-group p-value |
|---|---|---|---|
| from scratch v Cydoc educational tool | | | 2.35E-01 |
| Cydoc educational tool v Cydoc intake form | 0.05 | 0.0167 | 8.14E-05 |
| From scratch v Cydoc intake form | | | 1.20E-04 |

*Supplementary Table 4.* Average note quality scores as rated by a blinded, independent physician. Notes were scored on a scale from 1-5 (with 1 being the lowest and 5 being the highest) in clarity, comprehensibility, pertinent negatives, and pertinent positives.

| Average note quality ratings | | | | | | |
|---|---|---|---|---|---|---|
| | **from scratch** | | **educational tool** | | **intake form** | |
| | mean | standard deviation | mean | standard deviation | mean | standard deviation |
| **clarity** | 3.39 | 0.78 | 2.61 | 1.50 | 3.35 | 0.79 |
| **comprehensibility** | 3.06 | 0.80 | 2.83 | 1.38 | 3.24 | 0.97 |
| **pertinent positives** | 2.94 | 0.64 | 2.67 | 1.24 | 3.47 | 0.80 |
| **pertinent negatives** | 2.67 | 0.69 | 2.39 | 1.24 | 3.47 | 0.80 |

Supplementary *Table 5*. P-values for note quality score categories (clarity, comprehensibility, pertinent positives, pertinent negatives) across note writing method. P-values were calculated using the Wilcoxon paired signed rank test.

| Wilcoxon paired signed rank test | | | | |
|---|---|---|---|---|
| | p-values | | | |
| | Clarity | Comprehensibility | Pertinent Positives | Pertinent Negatives |
| From scratch compared to Note Writing Tool | 0.1422 | 0.9063 | 0.7263 | 0.8445 |
| From scratch compared to Patient Intake Form | 0.6891 | 0.6496 | 0.7299 | 0.4102 |
| Note-Writing Tool compared to Patient Intake Form | 0.1823 | 0.5936 | 0.2393 | 0.1549 |

Supplementary Table 6. *Average note quality rating by standardized patient (SP) scenario.*

| Average note quality ratings by SP scenario | | | | | | |
|---|---|---|---|---|---|---|
| | **back pain** | | **abdominal pain** | | **diabetes check-up** | |
| | mean | standard deviation | mean | standard deviation | mean | standard deviation |
| **clarity** | 2.89 | 1.45 | 2.76 | 1.35 | 2.71 | 1.36 |
| **comprehensibility** | 2.72 | 1.36 | 2.94 | 1.43 | 2.59 | 1.18 |
| **pertinent positives** | 2.72 | 1.32 | 2.88 | 1.27 | 2.76 | 1.20 |
| **pertinent negatives** | 2.61 | 1.33 | 2.65 | 1.32 | 2.59 | 1.18 |

*Supplementary Table 7. Note time-to-completion by order of scenario. The order of each scenario was altered for each participant to account for time exhaustion. The table below shows the average and standard deviation of time-to-completion for each note-writing method by scenario order.*

| Note time-to-completion by order of scenario | | | | | | |
|---|---|---|---|---|---|---|
| | **first scenario** | | **second scenario** | | **third scenario** | |
| | average | standard deviation | average | standard deviation | average | standard deviation |
| **from scratch** | 15.83 | 3.54 | 14.33 | 4.41 | 20.83 | 10.30 |
| **Cydoc educational tool** | 14.17 | 5.67 | 17.83 | 10.42 | 20.50 | 7.77 |
| **Cydoc intake form** | 7.40 | 2.70 | 6.00 | 3.58 | 4.00 | 2.00 |